\documentclass[]{rmaa}
\usepackage{natbib}

\usepackage{paralist}
\usepackage[T1]{fontenc}

\usepackage{psfrag,color}

\usepackage[latin1]{inputenc}
\usepackage{epsfig}
\usepackage{amsmath}
\usepackage{float}
\usepackage{graphicx}
\usepackage{subfig}
\usepackage{multirow}
\usepackage{array}
\usepackage{rotating}
\usepackage[normalem]{ulem}
\usepackage{wasysym}

\title{CCD PHOTOMETRY OF THE GLOBULAR CLUSTER $NGC$ $5897$\altaffilmark{1}}

\author{A. Ruelas-Mayorga\altaffilmark{2}, L. J. S\'anchez\altaffilmark{2}, E. Mac\'ias-Estrada \altaffilmark{2}, A. Nigoche-Netro\altaffilmark{3}}

\altaffiltext{1}{Based upon observations acquired at the Observatorio Astron\'omico Nacional on the Sierra San Pedro M\'artir (OAN-SPM), Baja California, M\'exico.}
\altaffiltext{2}{Instituto de
Astronom\'{\i}a, Universidad Nacional Aut\'onoma de M\'exico,
M\'exico D.F., M\'exico.}
\altaffiltext{3}{Instituto de Astronom{\'\i}a y Meteorolog{\'\i}a, Universidad
     de Guadalajara, Guadalajara, Jal. 44130, M\'exico.}

\fulladdresses{
 \item A. Ruelas-Mayorga, L.~J.~S\'anchez, E. Mac\'ias-Estrada: Instituto de Astronom\'{\i}a, Universidad
        Nacional Aut\'onoma de M\'exico, Apartado Postal 70-264, Cd.
        Universitaria 04510, M\'exico D.F., M\'exico.
        (\email{rarm,leonardo,emacias@astro.unam.mx}).
\item A. Nigoche-Netro: Instituto de Astronom{\'\i}a y Meteorolog{\'\i}a, Universidad
                                                                          de Guadalajara, Guadalajara, Jal. 44130, M\'exico. (\email{anigoche@gmail.com}).}

\shortauthor{Ruelas-Mayorga et al.}
\shorttitle{CCD Photometry of NGC 5897}

\abstract{We report CCD photometric observations of the globular cluster NGC 5897, in the Johnson system filters $B$, $V$, $R$, and $I$. With the values for these magnitudes we obtain various colour indices and produce several colour-magnitude diagrams.
We present eight colour-magnitude diagramas: $V$ vs $B-V$, $B$ vs $B-V$, $V$ vs $V-I$, $I$ vs $V-I$, $R$ vs $R-I$, $I$ vs $R-I$, $V$ vs $V-R$, and $R$ vs $V-R$. In all of these diagrams we can clearly see the Giant Branch, the Horizontal Branch and the beginning of the Main Sequence. To the left of the Main Sequence turn-off point we detect a somewhat large number of Blue Straggler stars. We determine the mean value of the visual magnitude of the $HB$ as $16.60 \pm 0.46$. This value is fainter than the value found by other authors.}

\resumen{Se reportan observaciones fotom\'etricas del c\'umulo globular NGC 5897 en los filtros del sistema de Johnson $B$, $V$, $R$ e $I$. Con los valores de estas magnitudes obtenemos diferentes \'indices de color y producimos varios diagramas color-magnitud. Presentamos ocho diagramas color-magnitud $V$ vs $B-V$, $B$ vs $B-V$, $V$ vs $V-I$, $I$ vs $V-I$, $R$ vs $R-I$, $I$ vs $R-I$, $V$ vs $V-R$, y $R$ vs $V-R$. En todos estos diagramas se distinguen claramente la Rama de las Gigantes, la Rama Horizontal y el comienzo de la Secuencia Principal. A la izquierda del punto de salida de la secuencia principal detectamos un n\'umero relativamente grande de estrellas `Blue Stragglers'. Determinamos el promedio de la magnitud visual de la HB como $16.60 \pm 0.46$. Este valor es m\'as d\'ebil que el valor encontrado por otros autores.}

\addkeyword{Galaxy -- globular clusters: individual NGC 5897 -- techniques: photometry}

\begin{document}

\maketitle

\section{INTRODUCTION}

\label{sec:intro}

The Globular Clusters are spherically symmetric stellar systems which may be found in all galaxies. They are very rich stellar systems that may be found in our Galaxy in remote regions of the galactic Halo and also close to the galactic centre. They contain hundreds of thousands of stars within a radius $20-50$ $pc$, having typical central densities between $10^{2}$ and $10^{4}$  $stars/pc^{3}$. Globular clusters are dynamically very stable and may live for a long time. They represent a remnant of a primordial stellar formation epoch and may be considered as proper galactic subsystems \citep{Ruelasetal2010}.

Many Milky Way globular clusters have been known for a long time (see the Messier ($1771$) catalogue). Presently we think there are approximately 150-200 in our Galaxy, although this figure does not include those clusters close to the Galactic Plane or very low surface brightness objects, (see \citet{monella} and \citet{Harris}).

Because Globular Clusters are very luminous systems they may be observed at very large distances. This fact  makes them fundamental in the study of Galactic Structure.

Some characteristics of Globular Clusters are:

\begin{itemize}

\item In general, the light that comes from these
objects originates in stars slightly cooler than our Sun.

\item Morphology. In general, they are slightly elliptical in shape. The average ratio between the minor and major axes of the apparent ellipse that they project on the sky is $b/a=0$.$73$, with only $5\%$ of them more elongated than $b/a=0$.$8$.

\item They appear to be exclusively stellar system in which no presence of gas or dust is detected \citep{galactic}.

\item The radial distribution of stars varies between clusters, and there are some that present a strong central concentration \citep{MIHALAS}.

\item The value of their integrated absolute magnitude $(M_{V})_{0}$ is usually found in the interval $-5>$($M_{V}$)$_{0}\gtrsim -10$
where the maximum of the distribution is found at $(M_{V})_{0}\approx -8$.$5$ and with a FWHM of $\sim \pm 1$ mag.

\item Their intrinsic colour takes values in the interval  $0$.$4\lesssim (B-V)_{0}\lesssim 0$.$8$ with a maximum at $(B-V)_{0}\approx 0$.$57$ \citep{MIHALAS}.

\item There are clusters with a large metallic deficiency, usually located in the Galactic Halo, up to those with abundances similar to that of the Sun (\citet{MIHALAS}, \citet{Harris2010}, \citet{Pfeffer2023}).

\item It is common to find clusters with metallic abundances in the interval $-2$.$2\lesssim \left[ Fe/H\right]\lesssim 0$.$0$.

\end{itemize}

Studies of Globular Clusters are important because they serve a variety of astronomical purposes, such as: determination of
the Galactic Centre position, as indicators of the galactic gravitational potential (i.e. \citet{Ishchenko2023} and references therein), studies of the evolution of low mass and low metallicity stars, and also of chemical evolution of galactic systems. Globular Clusters can also provide restrictions on galaxy formation (i.e. \citet{Strader2005}, \citet{Harris2013}, \citet{Beasley2020}, and \citet{Palma2023}).

$NGC$ $5893$ is a low concentration cluster $(c=log \ r_t/r\ c=1.19)$ \citet{Webbink1985}. There are in the literature several published Colour-Magnitude diagrams for this cluster (\citet{SandageKatem1968}, \citet{Sarajedini1992}, \citet{Ferraroetal1992} and \citet{Stetson2019}). \citet{SandageKatem1968} obtained $B$ and $V$ photometry down to $V \sim 17$ which was just able to include the Horizontal Branch (HB) of this cluster. They found that $V(HB)=16.20$, $E(B-V)=0.11 \pm 0.02$ and $(B-V)_{0,g}=0.78 \pm 0.03$ (the dereddened colour of the Giant Branch (GB) at the level of the HB), this allowed them to determine the metallicity of $NGC$ $5893$ to be between that of $M3$ and $M92$.  The HB of this cluster is made mainly by blue stars. \citet{Wehlau1990} found the average of the $V$ magnitude of the RR Lyrae star in this cluster to be $V(RR)=16.30$. \citet{Sarajedini1992} obtained CCD $B$ and $V$ photometry of this cluster down to $V \sim 22$. In this paper he reached the following conclusions:
\begin{itemize}
\item The HB of this cluster appears to be made mainly by blue stars with $V(HB)=16.35 \pm 0.15$.
\item The metallicity of this cluster is $\left[Fe/H\right]=-1.66 \pm 0.10$. He obtains this value from his adopted value for the $E(B-V)=0.07 \pm 0.04$, and the colour of the red giant branch $(RGB)$ at the level of the $HB$.
\item He finds that $NGC$ $5897$ is $\sim 2$ $Gyr$ older that the globular cluster $M3$.
\item The Luminosity Function of the $RGB$ shows an enhancement of the number of stars at the level of the $HB$.
\item The colour-magnitude $(CM)$ diagram of this cluster reveals  a large population of blue straggler stars.

\end{itemize}

This paper is organised as follows: in Section \ref{sec:obs} we present the observations and describe the reduction of the standard stars, in Section \ref{sec:clusterstars} the reduction of the photometry of the cluster stars is described, as well as the procedure for aligning and matching the different sections of the cluster which we observed. Section \ref{sec:photclusterstars} deals with the derivation of the Fiducial Lines and the introduction of the Colour-Magnitude diagrams, the following section (Section \ref{sec:CMdiagrams}) presents the different Colour-Magnitude diagrams which we derived. Section \ref{sec:metred} talks about the calculation of the metallicity and the reddening using the Sarajedini \& Layden method (see \citet{Sarajedini1994} and \citet{SarajediniLayden1997}). Section \ref{sec:modulus} derives the distance modulus to NGC 5897, and finally, in Section \ref{sec:conclusions} we present our conclusions.

\begin{table*}[!ht]
\caption{{}}
\label{tab:NGC5897tabla}\centering
\begin{tabular}{llc}
\hline\hline
\multicolumn{3}{c}{Data for the Globular Cluster NGC 5897} \\ \hline
\multicolumn{2}{l} {Right Ascension (2000)} & $15^{h}17^{m}24$.$5^{s}$ \\
\multicolumn{2}{l}{Declination (2000)} & $-21^{o} 00'37.0''$ \\
\multicolumn{2}{l}{Galactic Longitude} & $342$.$95$ \\
\multicolumn{2}{l}{Galactic Latitude} & $30$.$29$ \\
\multicolumn{2}{l}{Distance to the Sun(kpc)} & $12.5$ \\
\multicolumn{2}{l}{Tidal radius (arcmin) \citep{Webbink1985}} & $11.5$  \\
\multicolumn{2}{l}{Distance to the Galactic Centre (kpc)} & $7$.$4$ \\
\multicolumn{2}{l}{Reddening $E(B-V)$} & $0$.$09$ \\
\multicolumn{2}{l}{Horizontal Branch Magnitude (in $V$)} & $16$.$27$ \\
\multicolumn{2}{l}{Distance Modulus ($m-M$) (in $V$)} & $15.76$ \\
\multicolumn{2}{l}{Integrated $V$ Magnitude} & $8.53$ \\
\multicolumn{2}{l}{Absolute Visual Magnitude} & $-7.23$ \\
& $U-B$ & $0$.$08$ \\
Integrated Colour Indices & $B-V$ & $0$.$74$ \\
(no reddening correction) & $V-R$ & $0$.$50$ \\
& $V-I$ & $1$.$041$ \\
\multicolumn{2}{l}{Metallicity $[Fe/H]$} & $-1.90$ \\
\multicolumn{2}{l}{Integrated Spectral Type} & $F7$ \\

\multicolumn{2}{l}{Heliocentric Radial Velocity (km$/$s)} & $101.5$\\
\multicolumn{2}{l}{Central Concentration} & $0.86$ \\
\multicolumn{2}{l}{Ellipticity} & $0$.$08$ \\
\multicolumn{2}{l}{Nucleus Radius (arcmin)} & $1.40$ \\
\multicolumn{2}{l}{Mean Mass Radius (arcmin)} & $2.06$ \\
\multicolumn{2}{l}{Central Surface Brightness (in $V$) (magnitudes$/$%
arcsec)} & $20.53$ \\
\multicolumn{2}{l}{Logarithm of the luminous density at the centre ($L_{\odot}$/$pc^{3}$)} & $1.53$ \\ \hline
\end{tabular}
\end{table*}

\section{The Observations}
\label{sec:obs}

\begin{figure}[H]
\begin{center}
\includegraphics[width=10.0cm]{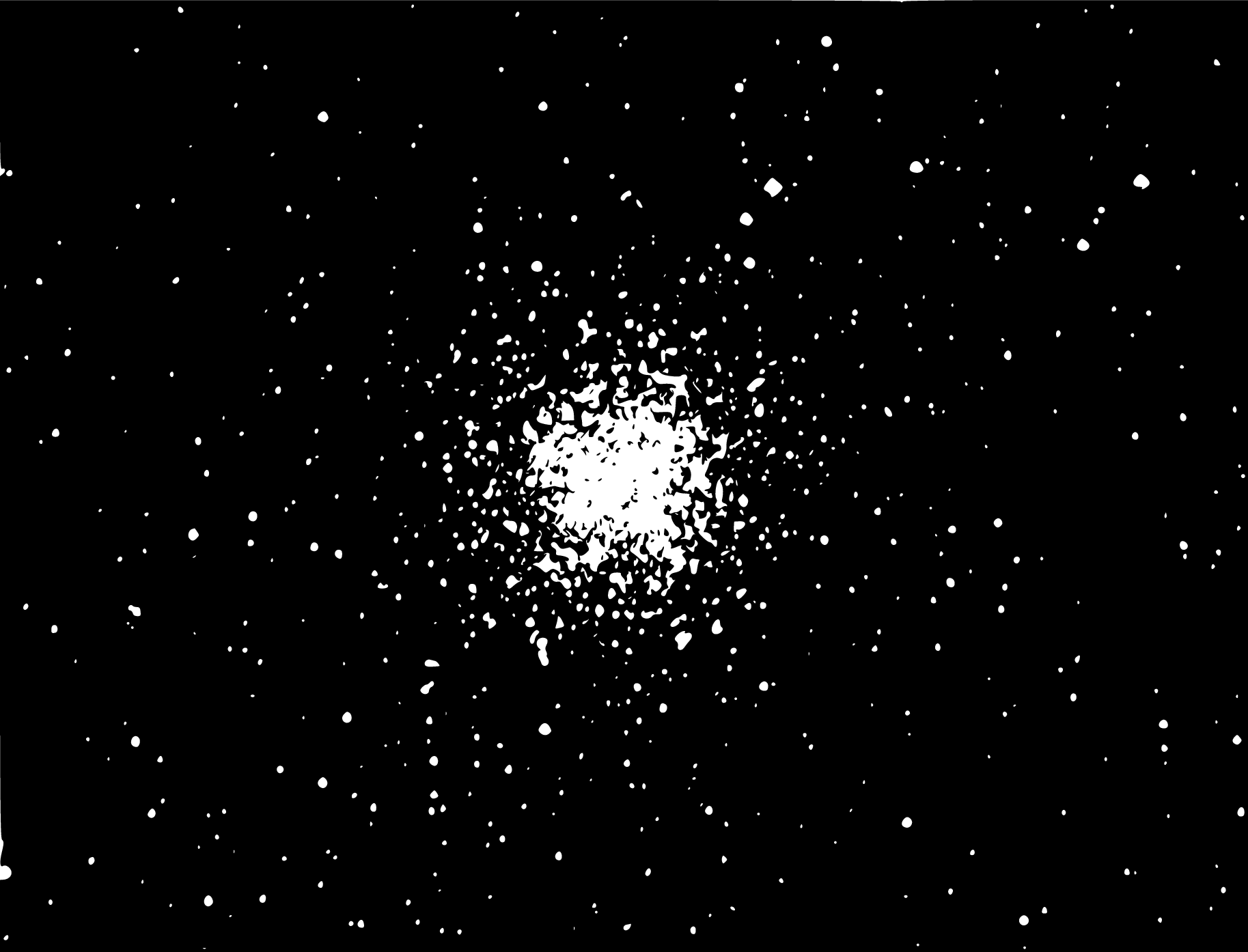}

\end{center}
\caption{The Globular Cluster NGC\ 5897. The image is $12.6\arcmin$ vertically (Taken from:
http://astrim.free.fr/ngc5897.htm/).} \label{fig:NGC5897}
\end{figure}

The globular cluster NGC 5897 is located on the constellation of Libra ($AR (2000:$ $15^{h}17^{m}24$.$5^{s}$,
$DEC (2000: -21^{o} 00'37.0'' )$, it contains several hundred thousands stars 
(see Figure \ref{fig:NGC5897}).
Its most important properties are listed in Table \ref{tab:NGC5897tabla}
\citep{Harris}.

We obtained the observations at the Observatorio Astron\'omico
Nacional in San Pedro M\'artir (OAN-SPM), Baja California during 2006, March 20-23 and 2007, March 14.

We utilised two different CCD cameras attached to the 1.5\,m telescope. The characteristics of these detectors are presented
in Table \ref{tab:detectordata}.


\begin{table*}[!htbp]
\caption{ }
\label{tab:detectordata}\centering%
\scalebox{1.0}[1.0]{
\begin{tabular}{ccc}
\hline\hline
\multicolumn{3}{c}{Characteristics of the Detectors Used in the Observations} \\ \hline
Characteristic & Thomson & Site1 \\ \hline
\multicolumn{1}{l}{Size (pixels)} & $2048\times 2048$ & $1024\times 1024$ \\
\multicolumn{1}{l}{Pixel Size ($\mu m$)} & $14\times 14$            & $24\times 24$ \\
\multicolumn{1}{l}{Quantum Efficiency}   & Maximum 65\% at 5000 \AA &  \\
\multicolumn{1}{l}{Reading noise ($e^{-}$) (gain } &  &  \\
mode $4$ binning $2\times 2$) & $5$.$3$ & $5$.$5$ (Direct Imaging) \\
\multicolumn{1}{l}{Dark Current ($e^{-}$/pix/h)} & $1$.$0$ a $-95$.$2$$^oC$ & $7$.$2$ approximately at $-80$$^oC$ \\
\multicolumn{1}{l}{Well Depth ($e^{-}$)} & $1$.$23\times 10^{5}$
(MPP Mode) &  \\
\multicolumn{1}{l}{Bias Level (gain } &  &  \\
mode $4$ binning $2\times 2$) & $384$ & $547$ (Direct Imaging) \\
\multicolumn{1}{l}{Gain ($e^{-}$) (Mode  $4$)} & $0$.$51$ & $1$.$27$ \\
\multicolumn{1}{l}{A/D Converter} & $16$ bits &  \\
\multicolumn{1}{l}{Linearity} &  $99\%$ &  $99$.$5\%$ \\
\multicolumn{1}{l}{Plate Scale ($\arcsec$/pixel)} & $0$.$147$
& $0$.$252$ \\ \hline
\end{tabular}}
\end{table*}

As stated above, the observations were collected during two observing runs, during which, we observed in a standard way
obtaining bias and flat field frames in each filter, plus obtaining multiple observations of standard star regions. In Tables \ref{tab:bitacora2006} and \ref{tab:bitacora2007} we present the logs for both observation seasons regarding the Globular Cluster NGC 5897.

\begin{table*}[!htbp]
\caption{ }
\label{tab:bitacora2006}\centering%
\scalebox{1.0}[1.0]{
\begin{tabular}{ccc}
\hline\hline
\multicolumn{3}{c}{Observation log for March 2006} \\ \hline
Object Observed & Filter & Number of Images \\ \hline
\multicolumn{1}{l}{Bias}             &              & $5$        \\
                                     &              &             \\
\multicolumn{1}{l}{Flat Field $B$}   & $B$          & $5$        \\
\multicolumn{1}{l}{Flat Field $V$}   & $V$          & $5$         \\
\multicolumn{1}{l}{Flat Field $R$}   & $R$          & $5$         \\
\multicolumn{1}{l}{Flat Field $I$}   & $I$          & $5$         \\
\multicolumn{1}{l}{Region $98$}      & $B$          & $4$        \\
\multicolumn{1}{l}{Region $98$}      & $V$          & $4$        \\
\multicolumn{1}{l}{Region $98$}      & $R$          & $4$        \\
\multicolumn{1}{l}{Region $98$}      & $I$          & $4$        \\
                                     &              &             \\
\multicolumn{1}{l}{Rubin $152$}      & $B$          & $4$        \\
\multicolumn{1}{l}{Rubin $152$}      & $V$          & $4$        \\
\multicolumn{1}{l}{Rubin $152$}      & $R$          & $4$        \\
\multicolumn{1}{l}{Rubin $152$}      & $I$          & $4$        \\
\multicolumn{1}{l}{Rubin $149$}      & $B$          & $4$        \\
\multicolumn{1}{l}{Rubin $149$}      & $V$          & $4$        \\
\multicolumn{1}{l}{Rubin $149$}      & $R$          & $4$        \\
\multicolumn{1}{l}{Rubin $149$}      & $I$          & $4$        \\
                                     &              &             \\
\multicolumn{1}{l}{PG $1323$  }      & $B$          & $4$        \\
\multicolumn{1}{l}{PG $1323$  }      & $V$          & $4$        \\
\multicolumn{1}{l}{PG $1323$  }      & $R$          & $4$        \\
\multicolumn{1}{l}{PG $1323$  }      & $I$          & $4$        \\
\multicolumn{1}{l}{PG $0942$  }      & $B$          & $4$        \\
\multicolumn{1}{l}{PG $0942$  }      & $V$          & $4$        \\
\multicolumn{1}{l}{PG $0942$  }      & $R$          & $4$        \\
\multicolumn{1}{l}{PG $0942$  }      & $I$          & $4$        \\                                                                  &              &             \\
\multicolumn{1}{l}{PG $1525$  }      & $B$          & $4$        \\
\multicolumn{1}{l}{PG $1525$  }      & $V$          & $4$        \\
\multicolumn{1}{l}{PG $1525$  }      & $R$          & $4$        \\
\multicolumn{1}{l}{PG $1525$  }      & $I$          & $4$        \\
\multicolumn{1}{l}{PG $1528$  }      & $B$          & $4$        \\
\multicolumn{1}{l}{PG $1528$  }      & $V$          & $4$        \\
\multicolumn{1}{l}{PG $1528$  }      & $R$          & $4$        \\
\multicolumn{1}{l}{PG $1528$  }      & $I$          & $4$        \\
                                     &              &             \\
\multicolumn{1}{l}{NGC $5897$ }      & $B$          & $10$        \\
\multicolumn{1}{l}{NGC $5897$ }      & $V$          & $10$        \\
\multicolumn{1}{l}{NGC $5897$ }      & $R$          & $10$        \\
\multicolumn{1}{l}{NGC $5897$ }      & $I$          & $10$       \\
\hline
\end{tabular}}
\end{table*}

\begin{table*}[!htbp]
\caption{ }
\label{tab:bitacora2007}\centering%
\scalebox{1.0}[1.0]{
\begin{tabular}{ccc}
\hline\hline
\multicolumn{3}{c}{Observation log for March 2007} \\ \hline
Object Observed & Filter & Number of Images \\ \hline
\multicolumn{1}{l}{Bias}                       &              & $10$        \\
                                               &              &             \\
\multicolumn{1}{l}{Flat Field $B$}             & $B$          & $3$        \\
\multicolumn{1}{l}{Flat Field $V$}             & $V$          & $3$         \\
\multicolumn{1}{l}{Flat Field $R$}             & $R$          & $3$         \\
\multicolumn{1}{l}{Flat Field $I$}             & $I$          & $3$         \\
                                               &              &             \\
\multicolumn{1}{l}{Rubin $152$}                & $B$          & $3$        \\
\multicolumn{1}{l}{Rubin $152$}                & $V$          & $3$        \\
\multicolumn{1}{l}{Rubin $152$}                & $R$          & $3$        \\
\multicolumn{1}{l}{Rubin $152$}                & $I$          & $3$        \\
                                               &              &             \\
\multicolumn{1}{l}{Rubin $149$}                & $B$          & $3$        \\
\multicolumn{1}{l}{Rubin $149$}                & $V$          & $3$        \\
\multicolumn{1}{l}{Rubin $149$}                & $R$          & $3$        \\
\multicolumn{1}{l}{Rubin $149$}                & $I$          & $3$        \\
                                               &              &             \\
\multicolumn{1}{l}{PG $1323$  }                & $B$          & $3$        \\
\multicolumn{1}{l}{PG $1323$  }                & $V$          & $3$        \\
\multicolumn{1}{l}{PG $1323$  }                & $R$          & $3$        \\
\multicolumn{1}{l}{PG $1323$  }                & $I$          & $3$        \\
                                               &              &             \\
\multicolumn{1}{l}{PG $0942$  }                & $B$          & $3$        \\
\multicolumn{1}{l}{PG $0942$  }                & $V$          & $3$        \\
\multicolumn{1}{l}{PG $0942$  }                & $R$          & $3$       \\
\multicolumn{1}{l}{PG $0942$  }                & $I$          & $3$        \\                                                              &              &             \\
\multicolumn{1}{l}{PG $1525$  }                & $B$          & $3$        \\
\multicolumn{1}{l}{PG $1525$  }                & $V$          & $3$        \\
\multicolumn{1}{l}{PG $1525$  }                & $R$          & $3$        \\
\multicolumn{1}{l}{PG $1525$  }                & $I$          & $3$        \\
                                               &              &             \\
\multicolumn{1}{l}{NGC $5897$ C, N, E, W}      & $B$          & $10$        \\
\multicolumn{1}{l}{NGC $5897$ C, N, E, W}      & $V$          & $10$        \\
\multicolumn{1}{l}{NGC $5897$ C, N, E, W}      & $R$          & $10$        \\
\multicolumn{1}{l}{NGC $5897$ C, N, E, W}      & $I$          & $10$       \\
\hline
\end{tabular}}
\end{table*}

\subsection{Reduction of standard stars}

In order to express the magnitude of the stars in the globular cluster in a standard system, we performed aperture photometry
of stars in some of the Landolt Standard Regions \citep{Landolt1992} listed in Tables \ref{tab:bitacora2006} and \ref{tab:bitacora2007}. The photometric observations of the standard stars was carried out using
the APT (Aperture Photometry Tool) \citep{Laheretal2012},\\ (see also https://www.aperturephotometry.org/about/).

The APT programme is a software that permits aperture photometry measurements of stellar images on a frame. The programme accepts images in the \textit{fits} format, so no transformation of our images to other formats was necessary. The standard regions which were measured had already been processed by removal of hot and cosmic ray pixels, and were bias and flat-field corrected.

We propose a set of transformation equations from the observed to the intrinsic photometric system that looks as follows:

 \begin{equation}
    B_{int}-B_{obs}=A_{B}*X+K_B*(B-V)_{obs}+C_{B}
    \label{eq:MC1}
 \end{equation}
 \begin{equation}
    V_{int}-V_{obs}=A_{V}*X+K_V*(B-V)_{obs}+C_{V}
    \label{eq:MC2}
 \end{equation}
 \begin{equation}
    R_{int}-R_{obs}=A_{R}*X+K_R*(R-I)_{obs}+C_{R}
    \label{eq:MC3}
 \end{equation}
 \begin{equation}
    I_{int}-I_{obs}=A_{I}*X+K_I*(R-I)_{obs}+C_{I}
    \label{eq:MC4}
 \end{equation}

    where the suffixes \textit{int} and \textit{obs} stand for \textit{intrinsic} and \textit{observed}, and $A$, $K$ and $C$ correspond to the negative of the atmospheric absorption coefficient, the colour term and the zero point term. These are the terms that we intend to calculate using the intrinsic values for the magnitudes given in \citet{Landolt1992}, and the observed values measured with APT. The equations are solved using the Least Squares procedure and the coefficients obtained are shown in Tables \ref{tab:CofFiltroB}, \ref{tab:CofFiltroV}, \ref{tab:CofFiltroR} and \ref{tab:CofFiltroI}.

   \begin{table}[!htbp]
   \caption{}
   \begin{center}
   \begin{tabular}{cccc}
   \hline
Standard System     &$\rm{A_B}$ & $\rm{K_B}$ & $\rm{C_B}$ \\
\hline
\hline

   21-22 March 2006 & -0.36496   & 0.19409    & 26.83682 \\
   22-23 March 2006 & -1.23555  & 0.17829    & 26.39759 \\
   23-24 March 2006 & -0.41733  & 0.42103    & 26.78603 \\
   12-13 March 2007 & 1.16919   & 0.10874    & 25.34031\\
   13-14 March 2007 & -0.30223  & 0.10583    & 27.47228 \\
   14-15 March 2007 & -0.37970  & 0.13717    & 27.53140 \\
   \hline
   \end{tabular}
   \label{tab:CofFiltroB}
   \end{center}
   \end{table}

   \begin{table}[!htbp]
   \caption{}
   \begin{center}
   \begin{tabular}{cccc}
   \hline
   Standard System     & $\rm{A_V}$  & $\rm{K_V}$    & $\rm{C_V}$  \\
   \hline
   \hline
   21-22 March 2006    & 0.94214     & 0.05596       & 25.03835 \\
   22-23 March 2006    & -1.67073    & 0.00563       & 28.07604 \\
   23-24 March 2006    & -0.46776    & 0.30053       & 27.38143  \\
   12-13 March 2007    &  0.12127    & -0.02352      & 27.58167 \\
   13-14 March 2007    & -0.25039    & -0.03984      & 27.76332 \\
   14-15 March 2007    & -0.32799    & -0.02824      & 27.86426 \\
   \hline
   \end{tabular}
   \label{tab:CofFiltroV}
   \end{center}
   \end{table}

   \begin{table}[!htbp]
   \caption{}
   \begin{center}
   \begin{tabular}{cccc}
   \hline
   Standard System     & $\rm{A_R}$  & $\rm{K_R}$  & $\rm{C_R}$    \\
   \hline
   \hline
   21-22 March 2006    & -----       & -----       & -----           \\
   22-23 March 2006    & -1.83264    & 0.13357     & 28.60749        \\
   23-24 March 2006    & -0.59665    & 0.50853     & 27.63788        \\
   12-13 March 2007    & 0.37819     & -0.02996    & 27.22330        \\
   13-14 March 2007    & -0.13806    & -0.04880    & 27.22305        \\
   14-15 March 2007    & -0.23895    & -0.05205    & 27.75572        \\
   \hline
   \end{tabular}
   \label{tab:CofFiltroR}
   \end{center}
   \end{table}

   \begin{table}[!htbp]
   \caption{}
   \begin{center}
   \begin{tabular}{cccc}
   \hline
   Standard System     & $\rm{A_I}$ & $\rm{K_I}$  & $\rm{C_I}$      \\
   \hline
   \hline
   21-22 March 2006    & -----      & -----       & -----           \\
   22-23 March 2006    & -2.25859   & 0.14873     & 29.11512         \\
   23-24 March 2006    & -0.56102   & 0.78455     & 27.07607         \\
   12-13 March 2007    & 1.14735    & 0.06524     & 25.31618         \\
   13-14 March 2007    & -0.02309   & 0.00138     & 26.50840         \\
   14-15 March 2007    & 0.11989    & -0.02645    & 26.57065         \\
   \hline
   \end{tabular}
   \label{tab:CofFiltroI}
   \end{center}
   \end{table}

\begin{figure}[H]
\begin{center}
\includegraphics[width=15.0cm,height=13.0cm]{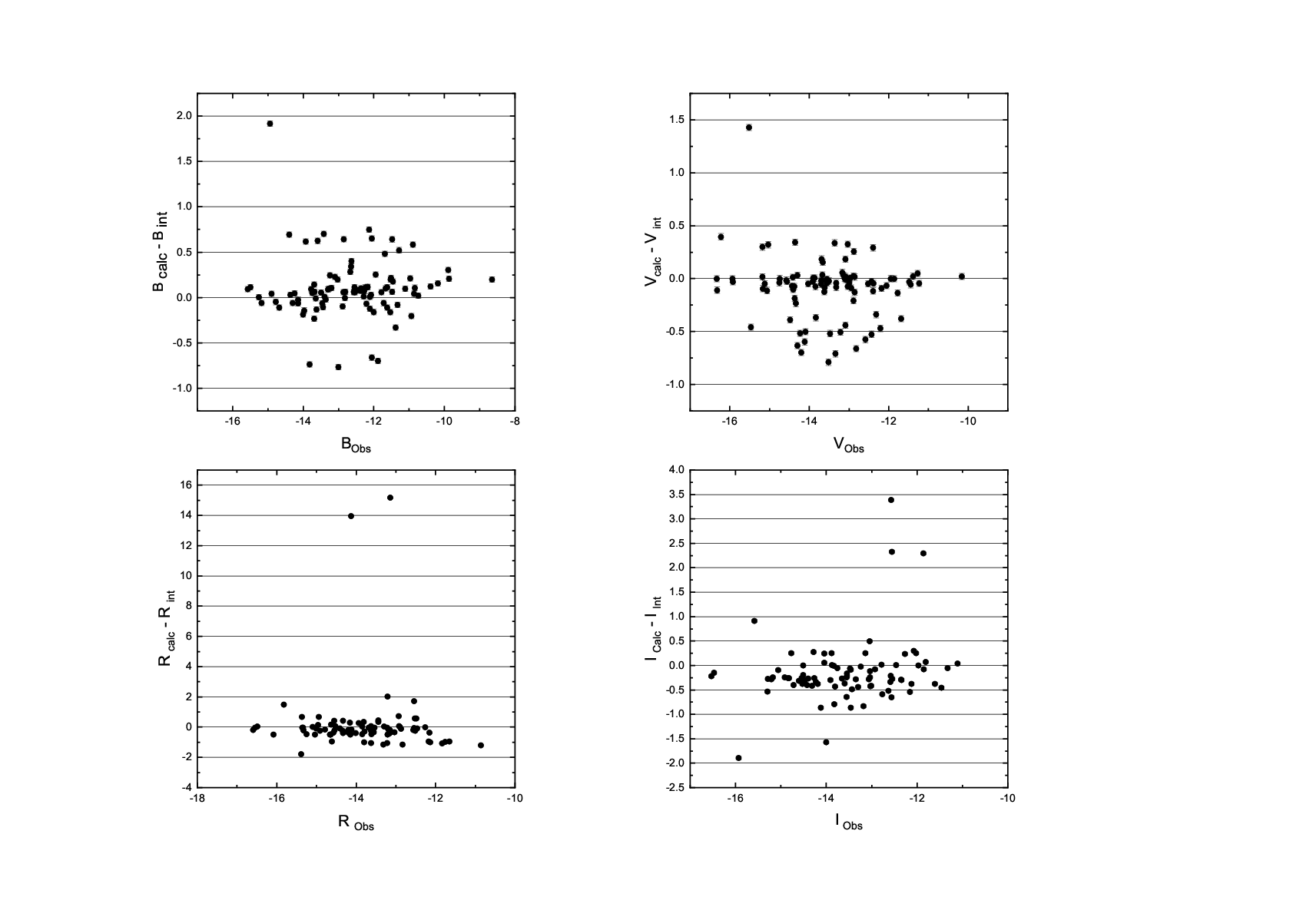}
\end{center}
\caption{Calculated minus intrinsic magnitude versus observed magnitude.} \label{fig:RESIDUALSTOGETHER}
\end{figure}

   In Figure \ref{fig:RESIDUALSTOGETHER} we plot, for magnitudes $B$, $V$, $R$ and $I$, on the vertical axis the difference between the intrinsic magnitude of the standard stars and their calculated values using the transformation equations (Equations \ref{eq:MC1}, \ref{eq:MC2}, \ref{eq:MC3} and \ref{eq:MC4}) given above and on the horizontal axis we plot the corresponding observed value for the magnitude. It is clear from these graphs that the calculated values minus the intrinsic values of the magnitudes for the standard stars are, in its majority, distributed around the zero value. However, there are two groups of stars that differ significantly from zero and appear displaced above and below the rest of the stars. These anomalous stars belong in general to the regions PG0942, Rubin 152, PG 1323 and Area 98. We have checked that the stars in these regions are well identified with our observation frames. However, we have been unable to find the cause of this discrepancy, therefore, these anomalous stars were eliminated from the rest of the analysis.

\begin{figure}[!htpb]
\begin{center}
\includegraphics[width=8.0cm,height=8.0cm]{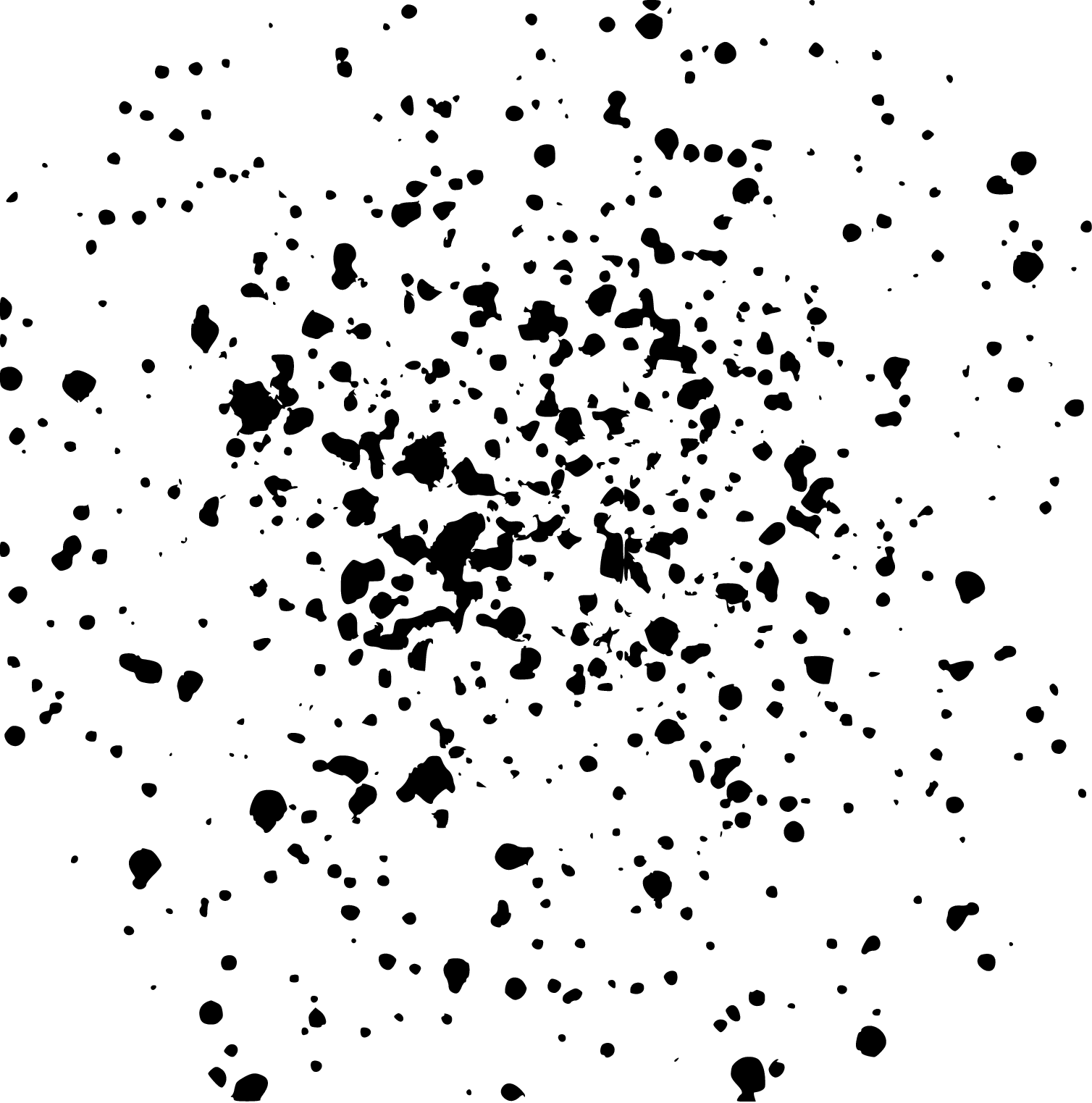}
\end{center}
\caption{Image of the  central zone of the Globular Cluster NGC 5897 in the R filter after the process of preparation.} \label{fig:RegioncentralR}
\end{figure}





The reduction of the data was done in a standard way, that is, removing dead and hot pixels produced by cosmic rays, and performing
bias subtraction, and flat field correction. This reduction process was achieved using the general purpose software: Image Reduction and Analysis Facility (IRAF). Once the fields
have been bias and flat field corrected we proceed to utilise the routine DAOPHOT to obtain the photometry of the
many starts present in the field using the Point Spread Function (PSF) technique \citep{Stetson1992}. As an example of an image
on which we apply the DAOPHOT technique see Figure \ref{fig:RegioncentralR}.

\begin{figure}[H]
\begin{center}
\includegraphics[width=15.0cm,height=13.0cm]{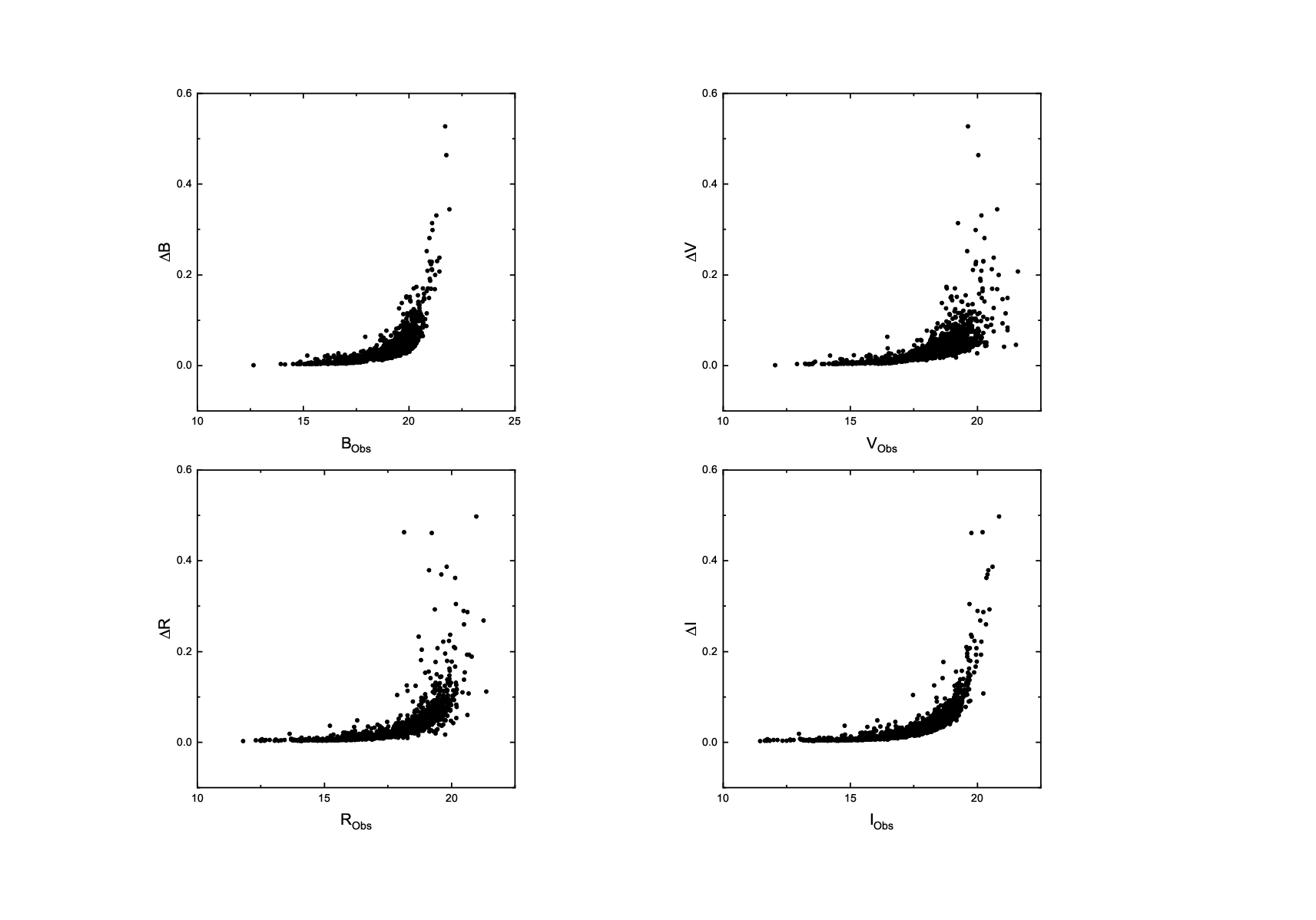}
\end{center}
\caption{Measurement error $\Delta Mag$ vs $Mag_{Obs}$ for $B$, $V$, $R$ and $I$.} \label{fig:DELTAMAGvsMAG}
\end{figure}

Figure \ref{fig:DELTAMAGvsMAG} show graphs of the measurement errors as function of the value of the observed magnitude for $B$, $V$, $R$ and $I$. It is clear that for brighter magnitudes the measurement errors are very small and begin to increase as we move to fainter magnitudes, becoming of the order of several tenths of magnitude by a magnitude value of $\sim 20$.

\section{Photometry of the Cluster Stars}
\label{sec:clusterstars}

The calculation of the different observed magnitudes of the stars in the cluster NGC 5897
was obtained by a standard application of the DAOPHOT subroutine present in IRAF to all the
observed frames in each filter, after having the frames bias-subtracted and flat fielded.

The magnitude catalogues produced by the subroutine ALLSTAR for the observations were combined
to obtained the observed colours for the stars in the globular cluster, so that the application of Equations \ref{eq:MC1}, \ref{eq:MC2}, \ref{eq:MC3} and \ref{eq:MC4} with the appropriate transformation coefficients was straightforward.

     \subsection{Colour-Magnitude diagrams}
     \label{DiagramsCM}
     In order to construct the colour-magnitude diagrams we utilised the photometric catalogues, in the four filters of interest $B$, $V$, $R$ and $I$, which were derived from the application of the DAOPHOT subroutines. The diagrams which we shall present are the following: ($ B $ vs $B-V$), ($V$ vs $B-V$), ($V$ vs $V-I$), ($I$ vs $V-I$), ($I$ vs $R-I$), ($R$ vs $R-I$), ($V$ vs $V-R$) and ($R$ vs $V-R$). To be able to use the derived photometric catalogues, it is necessary to effect a series of steps so that the frames in all filters are compatible, it was therefore necessary to align, group, and standardise them as well as eliminate those stars that might appear repeated.

     \subsection{Alignment}
     \label{Alignment}
     To obtain the colours of the star in NGC 5897 we need to pair the different magnitude catalogues. This procedure is not a trivial one since one star does not necessarily have the same coordinates in all the frames in which it appears. We have to remember that each filter observation was taken in 5 different frames (North, South, East, West and Centre) which have to be assembled into a grand image for the entire cluster, taking care to obtain the average of the intensity of those stars that appear on any two sections of the mosaic. Taking into consideration that we used one CCD in 2006 and a different one in 2007 the matching of the images required a positional transformation which consisted of a translation, a rotation and a stretching. Fortunately, the coefficients for rotation and stretching resulted in very small numbers, so our problem reduced itself to a simple translation of coordinates. We define as a primary reference frame the positions and the magnitudes of the stars in the central image observed for the 2007 observing season. Doing this, we ended up having photometric files for the four filters in the same positional and magnitude systems, which allowed the calculation of colours in an easy and straightforward way.

     \section{Photometry of the stars in NGC 5897, Colour Catalogues and Fiducial Lines}
     \label{sec:photclusterstars}
     The standard magnitudes for the stars in the cluster NGC 5897 were obtained by comparison with the published photometry results for this cluster published by \citet{Stetson2019}. A number of stars were identified in our and Stetson's observations, and a linear transformation between one set of observations and the other were proposed as follows:

      \begin{equation}
         \alpha = C_1 X_{pos} + C_2.
     \end{equation}
     Where $\alpha$ represents Right Ascension,  $C_1$ is the longitudinal stretching coefficient, $X_{pos}$ represents the $X$ coordinate value in our files and $C_2$ is the longitudinal coefficient of translation.\\
     Analogously,
     \begin{equation}
         \delta = C_3 Y_{pos} + C_4.
     \end{equation}
     Where $\delta$ represents the declination, $C_3$ is the transversal stretching coefficient, $Y_{pos}$ represents the $Y$ coordinate in our files and $C_4$ is the transversal coefficient of translation.\\

     The values we obtained for the transformation coefficients between our positions and those of Stetson's are as follows:

       \begin{itemize}
      \item $C_1=-9.60592 \times 10^{-5}$
      \item$C_2=229.39920$
      \item $C_3=8.76398 \times 10^{-5}$
      \item $C_4=-21.06008$
    \end{itemize}

    The magnitude transformation results simply on an additional displacement for the four filters. This procedure is applied to our four mosaics in filters $B$, $V$, $R$ and $I$ producing a final set of NGC 5897 stars in four filters on a standard system of position and magnitude.

    \subsection{Colour catalogues}
    \label{Colourcatalogues}
    We formed four colour catalogues as follows:
    \begin{itemize}

    \item $B-V$ catalogue 1656 stars
    \item $V-I$ catalogue 1935 stars
    \item $R-I$ catalogue 2026 stars
    \item $V-R$ catalogue 2587 stars

  \end{itemize}

    The number of stars in each catalogue is different because not all stars are equally detected in the different filters and therefore, some of them may be detected in one filter and not in another.

    \subsection{Fiducial lines (FL) and Colour-Magnitude diagrams (CMD)}
    \label{CMdiagramsFL}
    In this section we shall present the colour magnitude diagrams which we obtained from the photometry performed on the stars of the globular cluster NGC 5897. Each colour-magnitude diagram will be presented as a plot of magnitude versus colour, on which we have superimposed a series of lines which correspond to the Fiducial line for the Giant Branch and part of the Main Sequence. The Fiducial Line was obtained with the following procedure:
    \begin{itemize}

\item We calculate the maximum and minimum value of the magnitude interval.
\item We divide this interval in a number of magnitude bins.
\item Along each magnitude bin we find the number of stars in a number of colour bins. This allows us to find a distribution of the number of stars along a magnitude bin in bins of colour. We obtain a distribution histogram for the number of stars in each magnitude bin.
\item Each one of these histograms is fitted by a Gaussian function and we obtain for this function its maximum height $(A)$, its central value $(x_0)$ and its Standard deviation $(\sigma)$.
\item The Fiducial Line is formed from the points with coordinates $(x_0, m_0)$ where $m_0$ is the average magnitude in each magnitude bin.

\end{itemize}

Table \ref{tab:FLVvsBV} presents the FL derived by \citet{Sarajedini1992} (Columns 1 and 2), that derived by us in this paper (Columns 3, 4 and 5), an eyeball fit to our FL (Columns 6 and 7) and an eyeball fit to the \citet{Sarajedini1992} CMD made by us (Columns 8 and 9). Tables \ref{tab:FLBvsBV} and \ref{tab:FLVvsVI} give the colours, magnitudes and colour dispersion for the Fiducial lines, calculated with the procedure described above, for the B vs (B-V) and V vs (V-I) colour magnitude diagrams.

\begin{figure}[H]
       \centering
       \includegraphics[width=12cm]{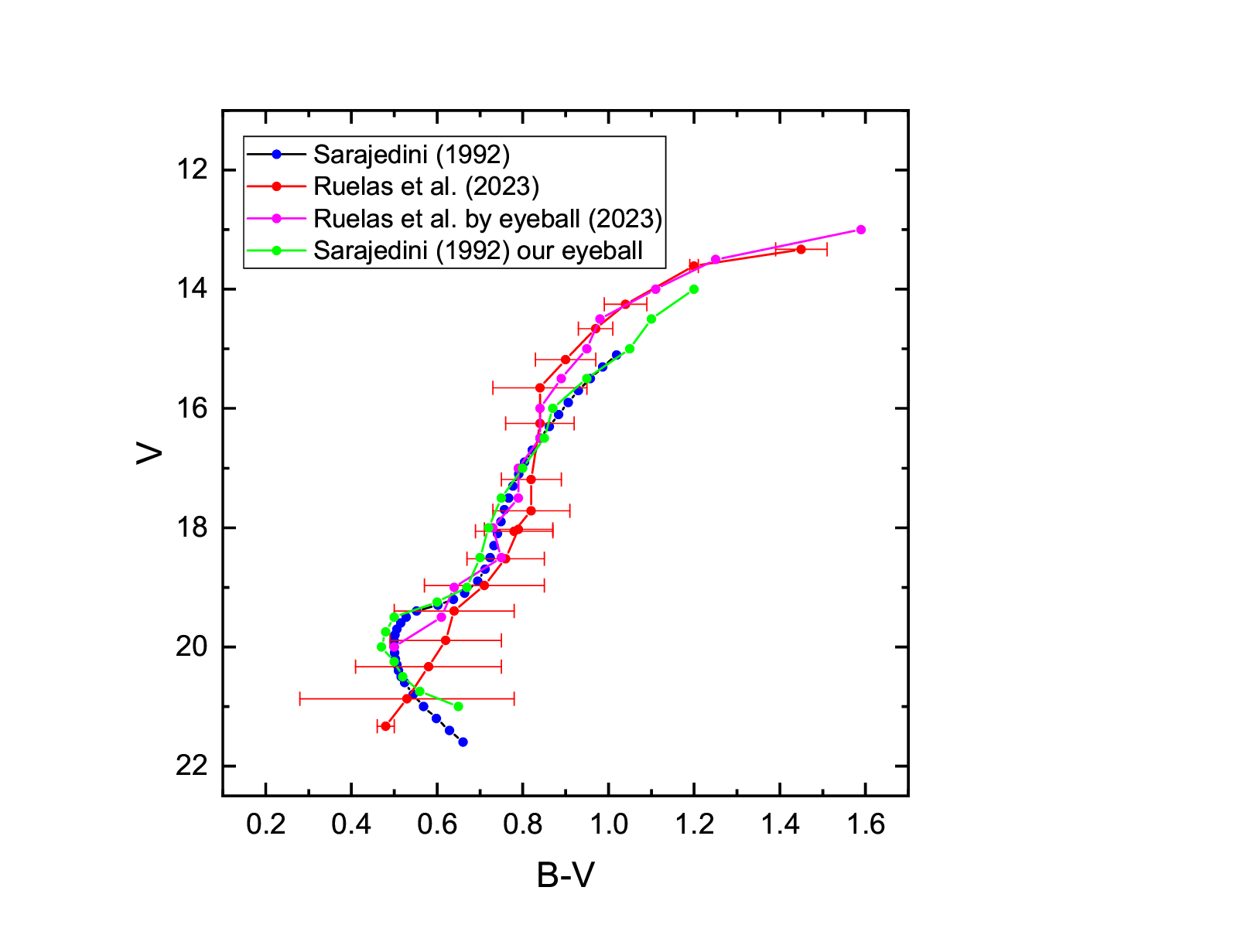}
       \caption{The blue dots show the Fiducial line derived by \citet{Sarajedini1992} from an eyeball fit to his CMD. The red dots show the FL derived in this paper following the procedure described herein (see text), error bars for these points are also shown. The pink dots show an eyeball fit to our FL and finally, the green dots show an eyeball fit to the Sarajedini CMD made by us.}
      \label{fig:COMPARISONFL}
     \end{figure}

Figure \ref{fig:COMPARISONFL} presents the FL derived by \citet{Sarajedini1992} (Blue) from an eyeball fitting to the points on his CMD and the FL we derived in this paper (Red) following the procedure described above. We have also included in this figure an eyeball fit to our FL (pink), as well as an eyeball fit made to the CM diagram of Sarajedini's made by us (green). The FL we derived using the Gaussian technique has errors in the $(B-V)$ colour that bring it in close agreement with the \citet{Sarajedini1992} Fiducial line. However, in the magnitude range $14.0 \leq V \leq 16.0$ our Fiducial line appears $\sim 0.1$ units bluer than that of Sarajedini's. Even if we consider the estimated error for the $(B-V)$ colour of our observations $(\sim 0.03)$ this cannot explain a shift of this magnitude. Fainter than $V \sim 16.0 $  and down to $V \sim 19.0 $ our FL and that of Sarajedini's coincide within the errors. Further down than $V \sim 19.0 $ there is again a discrepancy which, in this case, is simply due to the paucity of our data at these faint magnitudes.

The difference we observe between our FL and that of Sarajedini's at brighter magnitudes baffles us and we cannot find, at this point, a reasonable explanation for it. Our observations are quite incomplete at levels deeper than $V \sim 19$ that is why our FL shows neither the Turn-off point nor the beginning of the Main Sequence.


\begin{table}[H]
   \caption{Fiducial Lines for V vs (B-V).}
   \begin{center}
   \begin{tabular}{ccccccccc}
   \hline
  $ (B-V)_{Sa}$  & $V_{Sa}$  &   $(B-V)_{O}$     &   $V_{O}$  &   $\sigma_{O}$  & Eye $(B-V)_{O}$ & Eye $V_{O}$ & Eye $(B-V)_{Sa}$   & Eye  $V_{Sa}$    \\
   \hline
1.019	&	15.1	&	1.45	&	13.33	&	0.06	&	1.59	&	13	    &	1.2	    &	14	\\
0.987	&	15.3	&	1.2	    &	13.61	&	0.01	&	1.25	&	13.5	&	1.1	    &	14.5	\\
0.958	&	15.5	&	1.04	&	14.25	&	0.05	&	1.11	&	14	    &	1.05	&	15	\\
0.93	&	15.7	&	0.97	&	14.66	&	0.04	&	0.98	&	14.5	&	0.95	&	15.5	\\
0.906	&	15.9	&	0.9	    &	15.18	&	0.07	&	0.95	&	15	    &	0.87	&	16	\\
0.884	&	16.1	&	0.84	&	15.65	&	0.11	&	0.89	&	15.5	&	0.85	&	16.5	\\
0.862	&	16.3	&	0.84	&	16.25	&	0.08	&	0.84	&	16	    &	0.8	    &	17	\\
0.84	&	16.5	&	0.82	&	17.19	&	0.07	&	0.84	&	16.5	&	0.75	&	17.5	\\
0.822	&	16.7	&	0.82	&	17.72	&	0.09	&	0.79	&	17	    &	0.72	&	18	\\
0.804	&	16.9	&	0.78	&	18.06	&	0.09	&	0.79	&	17.5	&	0.7	    &	18.5	\\
0.791	&	17.1	&	0.79	&	18.03	&	0.08	&	0.73	&	18	    &	0.67	&	19	\\
0.777	&	17.3	&	0.76	&	18.52	&	0.09	&	0.75	&	18.5	&	0.6	    &	19.25	\\
0.767	&	17.5	&	0.71	&	18.97	&	0.14	&	0.64	&	19	    &	0.5	    &	19.5	\\
0.757	&	17.7	&	0.64	&	19.4	&	0.14	&	0.61	&	19.5	&	0.48	&	19.75	\\
0.749	&	17.9	&	0.62	&	19.89	&	0.13	&	0.5	    &	20	    &	0.47	&	20	\\
0.741	&	18.1	&	0.58	&	20.33	&	0.17	&		    &		    &	0.5	    &	20.25	\\
0.733	&	18.3	&	0.53	&	20.87	&	0.25	&		    &		    &	0.52	&	20.5	\\
0.724	&	18.5	&	0.48	&	21.33	&	0.02	&		    &		    &	0.56	&	20.75	\\
0.712	&	18.7	&		    &		&		&		&		    &	0.65	&	21	\\
0.695	&	18.9	&		    &		&		&		&		    &		    &		\\
0.664	&	19.1	&		    &		&		&		&		    &		    &		\\
0.638	&	19.2	&		    &		&		&		&		    &		    &		\\
0.602	&	19.3	&		    &		&		&		&		    &		    &		\\
0.552	&	19.4	&		    &		&		&		&		    &		    &		\\
0.528	&	19.5	&		    &		&		&		&		    &		    &		\\
0.515	&	19.6	&		    &		&		&		&		    &		    &		\\
0.506	&	19.7	&		    &		&		&		&		    &		    &		\\
0.502	&	19.8	&		    &		&		&		&		    &		    &		\\
0.5	    &	19.9	&		    &		&		&		&		    &		    &		\\
0.5	    &	20	    &		    &		&		&		&		    &		    &		\\
0.501	&	20.1	&		    &		&		&		&		    &		    &		\\
0.503	&	20.2	&		    &		&		&		&		    &		    &		\\
0.506	&	20.3	&		    &		&		&		&		    &		    &		\\
0.51	&	20.4	&		    &		&		&		&		    &		    &		\\
0.516	&	20.5	&		    &		&		&		&		    &		    &		\\
0.524	&	20.6	&		    &		&		&		&		    &		    &		\\
0.544	&	20.8	&		    &		&		&		&		    &		    &		\\
0.569	&	21	    &		    &		&		&		&		    &		    &		\\
0.598	&	21.2	&		    &		&		&		&		    &		    &		\\
0.629	&	21.4	&		    &		&		&		&		    &		    &		\\
0.661	&	21.6	&		    &		&		&		&		    &		     &		\\

   \hline
   \end{tabular}
   \label{tab:FLVvsBV}
   \end{center}
   \end{table}

\begin{table}[!htbp]
   \caption{Fiducial Line for B vs (B-V)}
   \begin{center}
   \begin{tabular}{ccc}
   \hline
  $ B-V$     & $B$      & $\sigma$      \\
   \hline
1.365	&	14.683	&	0.106	\\
1.05	&	15.238	&	0.068	\\
0.968	&	15.608	&	0.085	\\
0.895	&	16.078	&	0.098	\\
0.864	&	16.444	&	0.113	\\
0.838	&	16.969	&	0.081	\\
0.747	&	17.382	&	0.019	\\
0.814	&	17.925	&	0.073	\\
0.773	&	18.639	&	0.102	\\
0.737	&	18.888	&	0.055	\\
0.731	&	19.248	&	0.12	\\
0.718	&	19.592	&	0.253	\\
0.656	&	19.682	&	0.151	\\
0.611	&	19.993	&	0.143	\\
0.456	&	20.441	&	0.138	\\

   \hline
   \end{tabular}
   \label{tab:FLBvsBV}
   \end{center}
   \end{table}

   \begin{table}[!htbp]
   \caption{Fiducial Line for V vs (V-I).}
   \begin{center}
   \begin{tabular}{ccc}
   \hline
  $ V-I$     & $V$      & $\sigma$      \\
   \hline
1.544	&	13.325	&	0.092	\\
1.616	&	14.177	&	0.211	\\
1.314	&	14.254	&	0.179	\\
1.107	&	15.169	&	0.088	\\
1.062	&	15.576	&	0.114	\\
1.057	&	15.641	&	0.093	\\
1.062	&	16.109	&	0.058	\\
0.992	&	16.574	&	0.078	\\
0.971	&	17.08	&	0.08	\\
0.926	&	17.542	&	0.082	\\
0.894	&	18.026	&	0.08	\\
0.811	&	18.516	&	0.204	\\
0.775	&	18.977	&	0.179	\\
0.76	&	19.417	&	0.165	\\
0.806	&	19.883	&	0.167	\\
1.007	&	20.324	&	0.33	\\
1.355	&	20.858	&	0.026	\\
1.422	&	21.332	&	0.269	\\

   \hline
   \end{tabular}
   \label{tab:FLVvsVI}
   \end{center}
   \end{table}

\section{Colour-Magnitude diagrams}
\label{sec:CMdiagrams}

\begin{figure}[H]
       \centering
       \includegraphics[width=15cm,height=13cm]{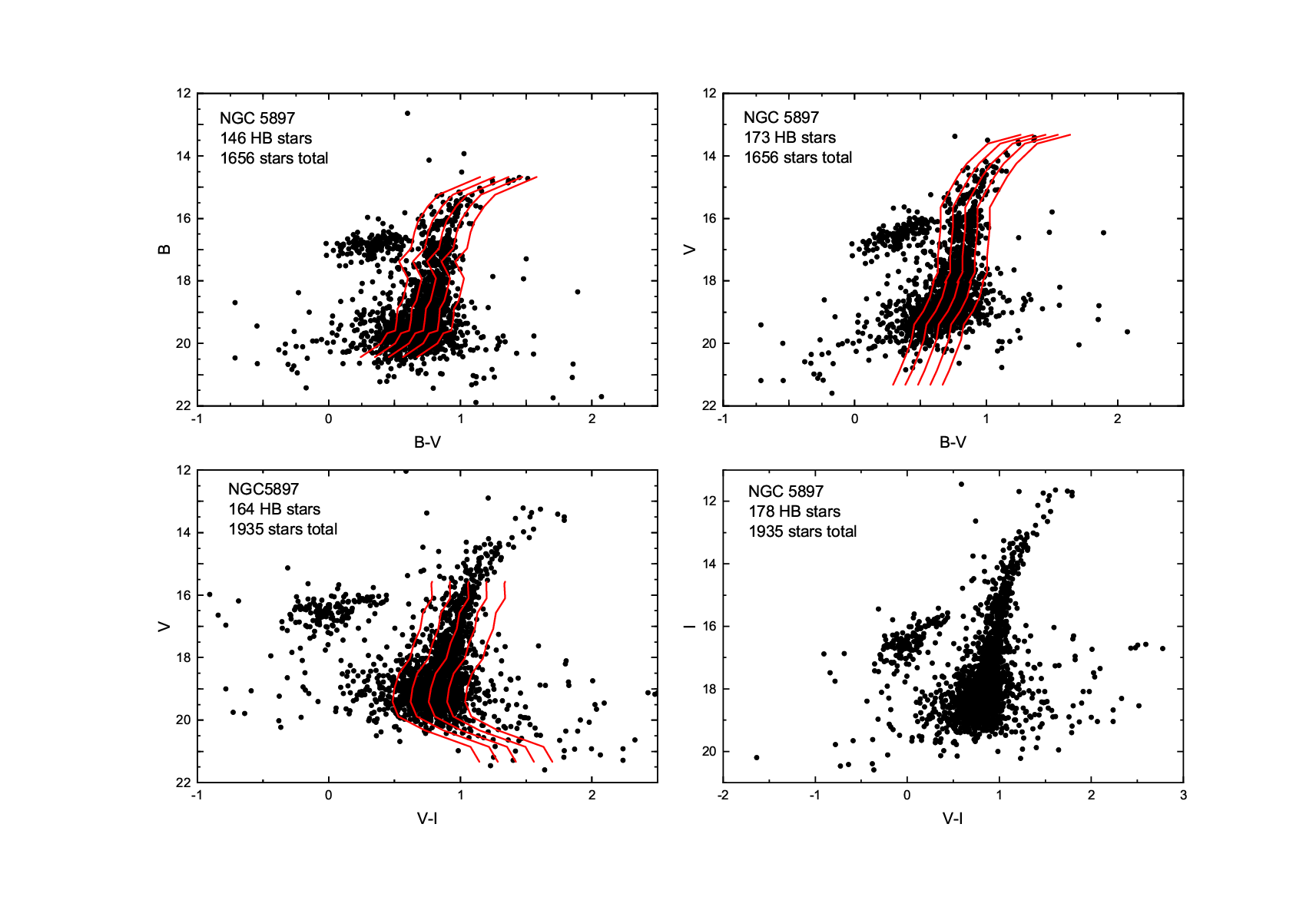}
       \caption{Colour-magnitude diagrams for $B$ vs $B-V$, $V$ vs $B-V$, $V$ vs $V-I$ and $I$ vs $V-I$. The central red line is the Fiducial Line (FL) (see text) and the other red lines indicate a $1\sigma$ and $2\sigma$ separation from the FL. The estimated errors are $\sim 0.02$ for the magnitudes and $\sim 0.03$ for the colours, smaller than the dots representing the stars.}
      \label{fig:COLMAGconFL}
     \end{figure}

    Figure \ref{fig:COLMAGconFL} shows the colour-magnitude diagrams $B$ vs $B-V$, $V$ vs $B-V$, $V$ vs $V-I$ and $I$ vs $V-I$.  The lines show the Fiducial Line (FL) obtained following the procedure described above and the FL displaced one and two Standard deviations. We can say that those stars that lie outside the $2\sigma$ limit and do not belong to the Horizontal Branch of the cluster have a very low probability $(\sim 5 \%)$ of belonging to the Globular Cluster.

\begin{figure}[H]
       \centering
       \includegraphics[width=15cm,height=13cm]{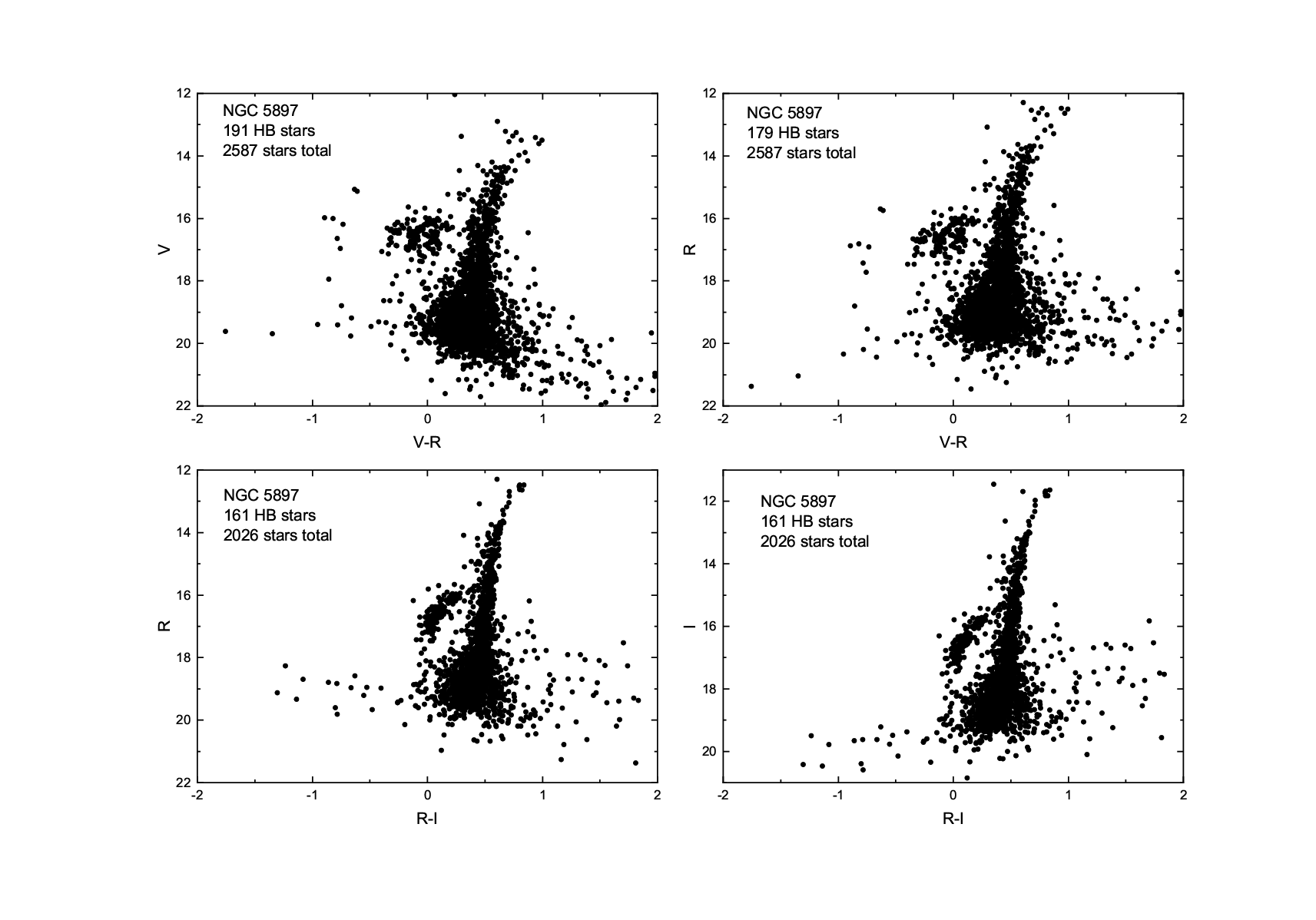}
       \caption{Colour-magnitude diagrams for $V$ vs $V-R$, $R$ vs $V-R$, $R$ vs $R-I$ and $I$ vs $R-I$. The estimated errors are $\sim 0.02$ for the magnitudes and $\sim 0.03$ for the colours, smaller than the dots representing the stars.}
       \label{fig:COLMAGsinFL}
     \end{figure}

     Figure \ref{fig:COLMAGsinFL} presents the colour-magnitude diagrams corresponding to the following combinations of magnitude and colour:$V$ vs $(V-R)$, $R$ vs $V-R$,  $R$ vs $(R-I)$ and $I$ vs $(R-I)$. On these diagrams we have not drawn the Fiducial Lines, but they can be easily obtained following the procedure described above.

     On all the colour-magnitude diagrams presented here we can clearly appreciate the presence of the Giant Branch (GB) which, on the $V$ vs $B-V$ diagram extends from $V \sim 13.5$ to $V \sim 20$ where the Main Sequence (MS) appears to begin. Our observations are not sufficiently deep to allow us a clear detection of the MS of this cluster. However, on the $V$ vs $V-R$ diagram we see stars that extend down to magnitude $V \sim 22$, where the Turn-off Point (TO) and the beginning of the MS are more clearly shown on this diagram. It is also clear that there is a number of stars that most probably do not belong to the globular cluster and that most certainly are foreground and to a lesser extent background stars. As mentioned above, the FLs obtained for three of the colour-magnitude diagrams permit us to eliminate with a high degree of certainty those stars which are located far away from the calculated fiducial line for the cluster. Below the HB and to the left (bluer colours) of the GB there is a large number of stars that could possibly be blue stragglers belonging to this cluster.

     The Giant Branch (GB) of this cluster is very well delineated on the $R$ vs $R-I$ and $I$ vs $R-I$ diagrams and it extends from $\sim 12$ to $\sim 18$ magnitudes.

     The presence of Horizontal Branch (HB) stars is obvious in all the diagrams. In those diagrams that include the $R$ and $I$ magnitudes, it is clear that the HB stars become fainter very quickly as we move towards bluer colours. In all our photometric catalogues we separated from the bulk of our observations the HB stars, and on each one of the colour-magnitude diagrams we explicitly state the number of HB stars belonging to this diagram.

     \subsection{The Horizontal Branch}
     \label{HB}
     The Horizontal Branch (HB) of globular cluster is an interesting region of their CM diagram because: i) it contains RR-Lyrae variable stars which have a fixed absolute magnitude (within certain restrictions). This fact is tremendously useful in the calculation of the distance of the Globular cluster in question. ii) It also presents groups of red and blue HB stars and the relative number of these red and blue stars is related to the Helium abundance of the stars in the cluster.

     In Table \ref{tab:averageHB} we give the mean value of the magnitude values for the stars in the Horizontal Branch for each filter which we observed. We shall later use these values in order to calculate a distance to $NGC$ $5897$. We also indicate the value of the median colour of the stars in the HB. It is clear that on all the CMD the median colour is located very much towards the blue edge of the HB, indicating that the majority of the stars in the HB branch are blue stars as mentioned in \citet{Sarajedini1992}.

   \begin{table}[!htbp]
   \caption{Mean value of the magnitude for the stars in the Horizontal Branch}
   \begin{center}
   \begin{tabular}{cccc}
   \hline
  CM   Diagram   & Mean HB Magnitude            & Standard        &  Median        \\
                 &                              & Deviation       &  colour         \\
\hline
$B$ vs $B-V$     &     $\overline{BHB}=16.84$   &   0.27          &  0.36          \\
$V$ vs $B-V$     &     $\overline{VHB}=16.60$   &   0.46          &  0.37          \\
$V$ vs $V-R$     &     $\overline{VHB}=16.57$   &   0.55          &  -0.05         \\
$R$ vs $V-R$     &     $\overline{RHB}=16.66$   &   0.51          &  -0.05         \\
$R$ vs $R-I$     &     $\overline{RHB}=16.63$   &   0.44          &  0.08          \\
$I$ vs $R-I$     &     $\overline{IHB}=16.57$   &   0.50          &  0.08          \\
$V$ vs $V-I$     &     $\overline{VHB}=16.60$   &   0.44          & -0.01          \\
$I$ vs $V-I$     &     $\overline{IHB}=16.58$   &   0.55          &  0.02          \\

   \hline
   \end{tabular}
   \label{tab:averageHB}
   \end{center}
   \end{table}

\section{Metallicity and Reddening}
\label{sec:metred}

The metallicity and the reddening of a globular cluster may be determined simultaneously by the Sarajedini method
(\citet{Sarajedini1994} and \citet{SarajediniLayden1997}), which takes into consideration the shape of the Red Giant Branch (RGB),
the observed magnitude $(V_{HB})$ of the Horizontal Branch (HB), the intrinsic $(B-V)$ colour of the RGB at the level of the HB; this value will be denoted herein as: $(B-V)_{0,g}$, and the difference in observed $V$ magnitude between the HB and the RGB at $(B-V)_{int}=1.2$
denoted by $\Delta V_{1.2}=V_{HB\ Obs}-V_{RGB\ Obs\ at \  1.2 \  int}$, where $V_{RGB\ Obs\ at \  1.2 \  int}$ means the observed $V$ magnitude of the RGB at an intrinsic $(B-V)$ colour equal to $1.2$. Defined this way, $\Delta V_{1.2} $ results in an intrinsically positive quantity.

We applied the Sarajedini method for the stars in the $V$ vs $(B-V)$ and the $V$ vs $(V-I)$ colour-magnitude diagrams and obtained values for the Metallicity $\left[Fe/H\right]$ and colour excesses $E(B-V)$ and $E(V-I)$. Unfortunately, these values were completely absurd, resulting in very high metallicities and negative colour excesses. We conducted a few numerical experiments and saw that the values of the metallicity and the colour excess are very sensitive to variations of the values of the fit coefficients. The coefficients we obtained for the FLs in the $V$ vs $(B-V)$ and $V$ vs $(V-I)$ diagrams had errors of the order $30 \  \%$ which might explain the troublesome results we obtained. We therefore decided to adopt the values given in the Harris catalogue (\citet{Harris}) $\left[Fe/H\right]=-1.90$ and $E(B-V)=0.09$.

The parameters discussed in this section could also be determined by means of isochrone fitting. There are in the literature several sets of isochrones which could be used (\citet{Girardi2002}, \citet{Spada2013}, \\ http://www.astro.yale.edu/demarque/yyiso.html, and references therein, \\ http://stev.oapd.inaf.it/cgi-bin/cmd and references therein. We shall report on this elsewhere.

\section{Distance Modulus}
\label{sec:modulus}

 Using the assumption that the average absolute magnitude of the HB is equal to the absolute magnitude of
the RR-Lyrae stars in the cluster, we calculate the distance to the cluster we study here  (for a justification of this assumption see, for example, \citet{Christy1966}, \citet{DemarqueMcClure1977} and \citet{Saio1977}).



For the RR-Lyrae stars, a linear relation between absolute magnitude and metallicity
of the form $M=a+b[Fe/H]$ is proposed in the astronomical literature. Determination of the constants $a$ and $b$ is achieved using the following methods: i)  statistical parallaxes, ii) the BBW moving atmosphere method, and iii) main sequence fitting (\citet{SandageTamman2006}).

In the following paragraphs we shall use different absolute magnitude-metallicity relations
from the literature
for RR-Lyraes, which combined with the metallicity and the apparent magnitude for the
HB of NGC 5897, will permit us to determine the value of its distance modulus.

 A compilation of statistical parallaxes of field RR-Lyrae stars has been presented by \citet{Wanetal1980} in a Catalogue of the
Shanghai Observatory. This compilation is summarised in
Table~3 of \citet{Reid1999}. There
is a value for ${\langle \mathrm{M_V}\rangle}_{\rm RR} = 0.83
\pm 0.23$ for $\langle{\rm [Fe/H]}\rangle$ values around $-$0.75
and another ${\langle \mathrm{M_V}\rangle}_{\rm RR} = 0.85 \pm
0.15$ for $\langle{\rm [Fe/H]}\rangle$ values around $-$1.56. A
linear extrapolation to the metallicity of NGC 5897 (${\rm [Fe/H]}=
-1.90)$ produces a value for

$${\langle \mathrm{M_V}\rangle}_{\rm NGC\ 5897} = 0.86 \, .$$

The extrapolation process poses a problem for this determination.

Using a combination of the infrared flux and the Baade-Wesselink analysis methods \citet{Fernley1989}, \citet{Fernley1990a}, \citet{Fernley1990b}, \citet{Skillenetal1989}, and \citet{Skillenetal1993} study 21 RR-Lyrae variable stars and obtain a mean relation for their absolute magnitude expressed as follows:
$$ {\langle \mathrm{M_V}\rangle}_{\rm RR} = (0.21\pm0.05){\rm [Fe/H]} + (1.04\pm0.10) \, ,$$
\noindent which for the metallicity value of our globular cluster produces a result of
$${\langle \mathrm{M_V}\rangle}_{\rm NGC\ 5897} = 0.64 \pm 0.20 \, .$$
\citet{Fernley1993} uses his near-IR Sandage Period-shift
Effect (SPSE) and a theoretical pulsation relation to derive the following relation:
$${\langle \mathrm{M_{V}}\rangle}_{\rm RR} = 0.19 {\rm [Fe/H]} + 0.84 \, ,$$
\noindent which applied to our cluster gives
$${\langle \mathrm{M_V}\rangle}_{\rm NGC5897} = 0.48\, .$$

\citet{McNamara1997} has reanalysed these same 29 stars making use of
more recent Kurucz model atmospheres and derives a steeper, more
luminous calibration given as follows:
$$ {\langle \mathrm{M_V}\rangle}_{\rm RR} = (0.29\pm0.05){\rm [Fe/H]} + (0.98\pm0.04) \, ,$$

The RR-Lyraes studied in the McNamara paper belong to
a metallicity interval from approximately -2.2 to 0.0. The metallicity
value for our globular cluster (-1.90) lies within this interval, making it reasonable to apply this relation to this cluster.
\noindent The value we obtain from this relation is:
$${\langle \mathrm{M_V}\rangle}_{\rm NGC\ 5897} = 0.43 \pm 0.14 \, .$$

\citet{Tsujimotoetal1998} have analysed data for 125  {\it Hipparcos}
RR Lyraes in the metallicity range $-2.49 <$ [Fe/H] $< 0.07$ using the maximum likelihood
technique proposed by Smith (1988). This technique allows simultaneous correction of the Malmquist and Lutz-Keller biases, allowing a full
use of negative and low-accuracy parallaxes. They derive the following relation:
$${\langle \mathrm{M_V\rangle}_{RR}} = (0.59 \pm 0.37) +
(0.20 \pm  0.63)({\rm [Fe/H]} +1.60) \, .$$
\noindent Given that $\mathrm{[Fe/H]_{NGC\ 5897}}=-1.90$ is contained within the studied
metallicity interval, applying this relation to the cluster studied in this paper produces
$${\langle \mathrm{M_V}\rangle}_{\rm NGC\ 5897} = 0.53 \pm 0.56 \, .$$

\citet{Arellanoetal2008a} and \citet{Arellanoetal2008b}  using the technique of Fourier
decomposition for the light curves of RR-Lyraes in several
globular clusters derive the following relation:

$${\langle \mathrm{M_{V}}\rangle}_{\rm RR} = +(0.18 \pm 0.03) {\rm [Fe/H]} + (0.85 \pm 0.05) \, .$$

This relation was obtained for a set of globular clusters contained within the metallicity interval
$-2.2 < \rm{[Fe/H]} < -1.2$ making it appropriate for the metallicity value (-1.90) we have for NGC 5897.

Applying this relation to our cluster we find

$${\langle \mathrm{M_V}\rangle}_{\rm NGC\ 5897} = 0.51 \pm 0.10 \, .$$

There are many different empirical and theoretical determinations
of the ${\langle \mathrm{M_V}\rangle}-{\rm [Fe/H]}$ relation for
RR-Lyrae stars, for ample discussions see \citet{Chaboyer1999}, \citet{CacciariClementini2003} and \citet{SandageTamman2006}. Determining
which one is the most appropriate for NGC 5897 is beyond the scope of
this paper, so we have decided to consider all of them for the calculation of the distance modulus of the
globular cluster studied in this paper.

From the data presented in
this paper we determine an apparent V
magnitude for the HB of NGC 5897 of $16.60 \pm 0.46$, which combined with the values for
the absolute magnitudes of the RR-Lyrae stars and the assumption that the HB and the RR-Lyraes have the same absolute magnitude yields
the distance modulus values presented in Table~\ref{tab:distmodul}.

\begin{table}
\caption{ }
\label{tab:distmodul}
\centering
\setlength{\tabcolsep}{0.5\tabcolsep}
\begin{tabular}{ccc}
\hline\hline
\multicolumn{3}{c}{DISTANCE MODULUS FOR NGC 5897}\\\hline

   From the Calibration given in     &     ${\langle \mathrm{M_V}\rangle}              $  &   $\rm{(m-M)_0}$  \\
                                     &                                                    & $m_{HB}-\langle \mathrm{M_V}\rangle-3.1E(B-V)$ \\\hline
                                     &                                                    &                     \\
\citet{Wanetal1980}                  &           $0.86 \pm 0.10$                          &   $15.66 \pm 0.10$  \\
\citet{Fernley1993}                  &           $0.48 \pm 0.10$                          &   $16.04 \pm 0.10$  \\
\citet{Skillenetal1993}              &           $0.64 \pm 0.20$                          &   $16.08 \pm 0.20$  \\
\citet{McNamara1997}                 &           $0.43 \pm 0.14$                          &   $16.17 \pm 0.14$  \\
\citet{Tsujimotoetal1998}            &           $0.53 \pm 0.56$                          &   $16.91 \pm 0.56$  \\
\citet{Arellanoetal2008a} and        &           $0.51 \pm 0.10$                          &   $16.01 \pm 0.10$  \\
\citet{Arellanoetal2008b}
                                     &                                                    &                     \\
\hline

\end{tabular}
\end{table}

A weighted average (by the inverse square of the errors) of these values results in an average distance modulus for
NGC 5897 of $15.96 \pm 0.64$ $(15500^{+5200}_{-3900} \ pc)$. The errors we encounter represent a $ \sim \pm 34 \%$ error in distance.
For the most part
the error in the distance modulus comes from the errors in the absolute magnitude versus
metallicity relations (see Table~\ref{tab:distmodul}, column 2), and not from
the errors in our photometry.

\citet{Benedictetal2002}, using the HST parallax for the prototype
RR-Lyrae star, determine an absolute magnitude for this star of
$M_v = 0.61 \pm 0.10$. If we assume that the HB of NGC 5897 has this
value for its absolute magnitude, then we obtain a distance modulus
of:
$$\rm{(m-M)_0}=(16.60 \pm 0.46)-(0.61 \pm 0.10)$$
$$-3.1 \times (0.09 \pm 0.10)=15.71 \pm 0.87$$

which agrees, within the errors, with previous determinations. This value
of the distance modulus produces a distance of $13800^{+6800}_{-4600}$ pc
to NGC 5897 with an error of $ \sim \pm 49 \%$.

We adopt as our best determination for the NGC 5897 distance modulus the average of the values
obtained with the \citet{Fernley1993} ($16.04 \pm 0.10$), and the \citet{Arellanoetal2008a} and \citet{Arellanoetal2008b} ($16.01 \pm 0.10$) calibrations
due to the fact that these calibrations presents the smallest errors. This average results in $16.02 \pm 0.14$.

\section{Conclusions}

\label{sec:conclusions}

In this paper we present $B$, $V$, $R$ and $I$ CCD photometry for the globular cluster NGC 5897. We obtained aperture photometry for a number of standard stars in the \citet{Landolt1992} regions, and then compared our observed stars with the magnitudes published by \citet{Stetson1992}. After aligning and matching the different sections of the globular cluster, we were able to produce magnitude catalogues for all the filters (a sample of these catalogues is presented in Appendix A). We formed eight colour magnitude (CM) diagrams (see Figures \ref{fig:COLMAGconFL} and \ref{fig:COLMAGsinFL}). In all these CM diagrams we can clearly see the Red Giant Branch (RGB), the Horizontal Branch (HB) and the beginning of the Main Sequence (MS). Towards bluer colours from the MS Turn-off point, we see a somewhat large number of stars that we identify with Blue Stragglers stars in this cluster.  We tried to calculate the metallicity and the reddening of this cluster making use of the Sarajedini-Layden method (see \citet{Sarajedini1994} and \citet{SarajediniLayden1997}), but we were unable to do so. We calculated the distance modulus to this cluster, and it resulted in $16.02 \pm 0.14$ which corresponds to a distance of $16.0 \pm 1.0$ $kpc$.

\section{Acknowledgements}

We would like to thank the Instituto de Astronom\'ia at Universidad Nacional Aut\'onoma de M\'exico (IAUNAM) and the Instituto de Astronom\'ia y Metereolog\'ia at Universidad de Guadalajara (IAMUdeG) for providing a congenial and stimulating atmosphere in which to work. We also thank the computing staff at both institutions for being always available and ready to help with random problems with our computing equipment, which arise when one least expects them. We would also like to thank Juan Carlos Yustis for help with the production of the figures in this paper.  We also thank Direcci\'on General de Asuntos del Personal Acad\'emico, DGAPA at UNAM for financial support under projects PAPIIT IN103813, IN102517 and IN102617. The help and valuable suggestions provided by an anonymous referee are gratefully acknowledged.

 \bibliography{librero1}

 \section{Appendix A}
 \label{appendixA}

 In Tables \ref{tab:BandVmags} and \ref{tab:RandImags} we present samples of the magnitude catalogues for this paper. The complete catalogues are available upon request.

\begin{table*}[!htbp]
\caption{ } \label{tab:BandVmags} \centering
\setlength{\tabcolsep}{0.5\tabcolsep}
\begin{tabular}{cccc}
\hline\hline
\multicolumn{4}{c}{B and V MAGNITUDES STARS IN NGC 5897}\\\hline
x (pixels)	        &	y (pixels)	        &	Mag B	&	Mag V	\\\hline
-277.6176	&	547.1518	&	19.4144	&	18.689125	\\
-260.9676	&	469.6858	&	19.8054	&	19.129125	\\
-242.1806	&	360.0848	&	16.6974	&	16.453125	\\
-225.8216	&	444.9578	&	18.4854	&	17.707125	\\
-217.0566	&	868.2468	&	19.0594	&	18.111125	\\
-211.4806	&	404.3458	&	19.1194	&	18.453125	\\
-210.1306	&	444.8878	&	19.9204	&	19.279125	\\
-209.5756	&	285.4678	&	18.1414	&	17.317125	\\
-206.3026	&	784.5288	&	18.7414	&	18.202125	\\
-203.2576	&	920.8858	&	16.9184	&	16.716125	\\
            &               &           &               \\
-175.0646	&	374.1218	&	18.1674	&	17.363125	\\
-174.9546	&	461.2768	&	20.1134	&	19.484125	\\
-167.2466	&	948.8658	&	19.0664	&	18.332125	\\
-165.9416	&	702.2378	&	15.4514	&	14.378125	\\
-155.0016	&	218.2148	&	19.2874	&	18.574125	\\
-154.4836	&	140.9378	&	19.7204	&	19.065125	\\
-147.7006	&	607.6628	&	16.9474	&	16.747125	\\
-128.4606	&	279.8978	&	16.5364	&	16.210125	\\
-128.0146	&	868.9678	&	16.7764	&	16.562125	\\
-120.0366	&	848.4458	&	18.8774	&	18.148125	\\
            &               &           &               \\
-96.1348	&	422.0696	&	18.9894	&	18.215125	\\
-88.0116	&	218.9878	&	16.1654	&	15.098125	\\
-87.8078	&	424.5506	&	20.0354	&	19.468125	\\
-85.0406	&	473.5338	&	19.1244	&	18.368125	\\
-83.6186	&	457.5028	&	18.6764	&	17.788125	\\
-80.8708	&	206.5076	&	20.0504	&	19.696125	\\
-78.9268	&	-112.1904	&	20.1914	&	19.493125	\\
-77.6868	&	463.3066	&	20.2104	&	19.453125	\\
-77.6686	&	746.1548	&	18.2714	&	17.494125	\\
-76.6066	&	1057.1788	&	20.0354	&	19.186125	\\

\hline
\end{tabular}
\end{table*}

\begin{table*}[!htbp]
\caption{ } \label{tab:RandImags} \centering
\setlength{\tabcolsep}{0.5\tabcolsep}
\begin{tabular}{cccc}
\hline\hline
\multicolumn{4}{c}{R and I MAGNITUDES STARS IN NGC 5897}\\\hline
x (pixels)	        &	y (pixels)	        &	Mag R	&	Mag I	\\\hline
-287.2978	&	628.6134	&	16.95635	&	16.55932	\\
-275.8288	&	546.6794	&	18.23635	&	17.89732	\\
-253.0268	&	552.9184	&	18.94435	&	18.63932	\\
-241.8808	&	359.8734	&	16.36735	&	16.32632	\\
-236.2428	&	805.0314	&	18.63835	&	18.35832	\\
-225.3138	&	444.7094	&	17.27135	&	16.85532	\\
-216.0218	&	867.7194	&	17.56735	&	17.18532	\\
-210.9128	&	405.2264	&	18.09135	&	17.74332	\\
-209.2018	&	285.3604	&	16.86735	&	16.47632	\\
-208.7448	&	670.3394	&	17.18435	&	16.76832	\\
            &               &               &               \\
-208.3818	&	945.1264	&	18.31335	&	17.93632	\\
-205.1298	&	784.2834	&	17.87935	&	17.64832	\\
-202.7348	&	919.9534	&	16.68035	&	16.68832	\\
-187.5238	&	817.3414	&	18.58435	&	18.19432	\\
-184.7668	&	573.4634	&	19.30235	&	18.93232	\\
-174.8078	&	373.8214	&	16.89735	&	16.48132	\\
-174.5678	&	461.5674	&	19.08335	&	18.96432	\\
-173.3298	&	625.7874	&	18.87335	&	18.70232	\\
-166.6518	&	948.2634	&	17.86835	&	17.49232	\\
-165.4968	&	701.9594	&	13.74835	&	13.19532	\\
            &               &               &               \\
-154.6808	&	364.3284	&	18.43835	&	18.01332	\\
-154.1348	&	218.2784	&	18.11335	&	17.87332	\\
-153.8348	&	353.4634	&	18.91635	&	18.60632	\\
-153.8298	&	141.3604	&	18.79335	&	19.65532	\\
-147.3108	&	607.1884	&	16.71035	&	16.72932	\\
-146.6818	&	407.2014	&	18.29635	&	17.68232	\\
-140.7538	&	880.8234	&	17.97135	&	17.56732	\\
-138.8768	&	586.5374	&	18.46135	&	17.96932	\\
-132.2968	&	412.4374	&	18.94335	&	18.67032	\\
-128.3178	&	279.7424	&	16.07935	&	15.98132	\\

\hline
\end{tabular}
\end{table*}

 \end{document}